\documentclass{elsart}
\usepackage{natbib}
\usepackage{psfig}

\def\lsim{\mathrel{\rlap{\lower 4pt \hbox{\hskip 1pt $\sim$}}\raise 1pt \hbox
        {$<$}}}
\def\gsim{\mathrel{\rlap{\lower 4pt \hbox{\hskip 1pt $\sim$}}\raise 1pt \hbox
        {$>$}}}

\begin{document}
\runauthor{Mineshige et al.}
\begin{frontmatter}
\title{Slim Disk Model for Narrow-Line Seyfert 1 Galaxies}
\author[Kyoto]{Shin Mineshige, Toshihiro Kawaguchi,}
\author[Osaka]{Mitsuru Takeuchi}
\address[Kyoto]{Department of
Astronomy, Kyoto University, Sakyo-ku, Kyoto 606-8502, Japan}
\address[Osaka]{Astronomical Institute,
Osaka Kyoiku University, Kashiwara 582-8582, Japan}

\begin{abstract}
We argue that both the extreme soft X-ray excess
and the large-amplitude variability of
narrow-line Seyfert 1 galaxies (NLS1s)
can be explained in the framework of the slim disk model.
When the disk luminosity approaches the Eddington luminosity,
the disk becomes a slim disk, 
exhibiting a multi-color blackbody spectrum 
with a maximum temperature, $T_{\rm bb}$,
of $\sim 0.2 (M/10^5 M_\odot)^{-1/4}$keV, and
size of the X-ray emitting region, $r_{\rm bb}$, of 
$\sim r_{\rm S}$ (the Schwarzschild radius).
Furthermore, magnetic energy can be amplified up to a level
exceeding radiation energy emitted from the disk,
causing substantial variability in X-rays
by consecutive magnetic flares.
\end{abstract}
\end{frontmatter}

\section{Introduction}

It has been recently established that
NLS1s are characterized by extreme soft excesses and
extreme variability \cite{Lei}, although
the origin still remain a puzzle.
Since these features are quite reminiscent of those of
Galactic black hole candidates during the very high state,
i.e.\ the state in which the luminosity is comparable to the Eddington
luminosity, $L_{\rm E}$,
it is natural to assume that NLS1s have a systematically
large disk luminosity, $L$ (e.g. \cite{Bol}).  In such a case, 
the disk is known to become a slim disk \cite{Abra}.
What, then, is the observational signature of the slim disk?

\section{Spectra of slim disks}

It will be useful to quickly review disk theory.
The standard disk model was constructed using energy balance
between viscous heating and radiative cooling with
advective energy transport assumed negligible.
Hence, accretion energy can be efficiently converted to radiation
energy.  This contrasts with the case of the advection-dominated flow, 
in which viscous heating is balanced by advective cooling;
that is, accretion energy is stored
in the accreting gas or trapped photons within the flow and is finally
swallowed by the black hole.  The slim disk corresponds to
the optically thick, advection-dominated flow.

\begin{figure}
\centerline
{\psfig{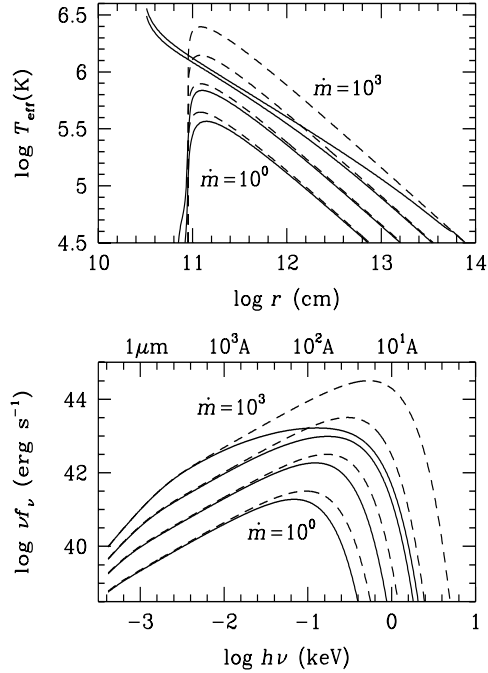}}
\caption
{
The temperature profiles (upper) and the emergent
spectra (lower) of accretion disks for various mass-flow rates;
from the bottom,
$\dot m[\equiv L/(L_{\rm E}/c^2)]=10^0, 10^1, 10^2$, and $10^3$.
The black-hole mass is fixed at $M=10^5M_\odot$ 
($r_{\rm S}\sim 10^{10.5}$cm).
As $\dot m$ increases, the regions with a flat temperature profile
expand to larger radii
and the disk spectra become flatter around the peak.
For comparison, we plot the temperature profiles expected from
the standard-disk relation (upper)
and their emergent spectra (lower) using dashed lines (see \cite{Mine}).}
\end{figure}

The slim disk exhibits rather unique temperature profiles.
We calculate the disk structure, taking into account the advection,
and plot in Figure 1 the effective temperature distribution (upper)
and spectra (lower), together with those of
standard disks (see \cite{Mine} for details).
When the mass-flow rate is relatively small, 
$\dot m\equiv {\dot M}/(L_{\rm E}/c^2) = 10$, the disk resides in
the standard-disk regime. When it increases up to $\sim 100$,
the two disk models give distinctly different results.
First, the slim disk exhibits a flatter temperature profile,
$T_{\rm eff}\propto r^{-1/2}$, in contrast with
$T_{\rm eff}\propto r^{-3/4}$ for the standard disk.
Second, substantial radiation arises from inside
the marginally stable last circular orbit at $r=3r_{\rm S}$.
This reflects the fact that, although the accretion velocity
is large, near the speed of light close to a black hole,
substantial material exists there because of the large $\dot M$.
Third, compared with the standard disk, the slim disk
produces higher energy photons, but the number of photons is fewer.
If we fit the spectra with a blackbody,
we find small sizes for the emitting region, $r_{\rm bb}\lsim r_{\rm S}$,
and high temperatures, $T_{\rm bb}\gsim 0.2 (M/10^5M_\odot)$ keV.

\begin{figure}[t]
\centerline
{\psfig{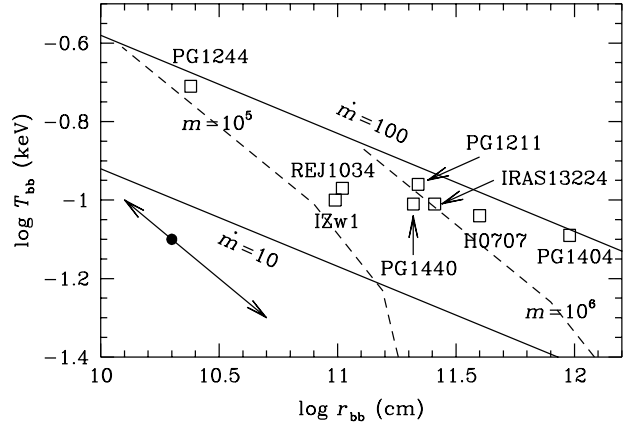}}
\caption{
The $r_{\rm bb}$--$T_{\rm bb}$ diagram of the observational
data from NLS1s.
The loci of $\dot m=10$ and 100 (solid lines)
and those of $m (\equiv M/M_\odot)= 10^5$ and $10^6$ (dotted lines)
obtained from the slim disk model are also shown.
The arrow pointed in the  upper-left (or lower-right) direction
from the filled circle in the lower-left corner indicates
the direction of the correction for that point to remove Compton and 
general relativistic effects
in the case of a face-on (nearly edge-on) disk (see \cite{Mine}).
}
\end{figure}

In Figure 2 we plot the theoretical values of $r_{\rm bb}$ and $T_{\rm bb}$
as functions of $\dot m$ and $m(\equiv M/M_\odot)$,
together with the ASCA data for NLS1s (see \cite{Mine} for details).
 From these plots, we can safely conclude
that all NLS1s plotted here fall into the regions above
$\dot m=10$.  This justifies our assumption
of large $L/L_{\rm E}$ in NLS1s.
Also, the derived black-hole masses are relatively small.

\section{Variability of Slim Disks}
A further issue is the variability.
We pay special attention to
the following characteristics of the variability:
(i) fluctuation light curves seem to be composed of
numerous flares (or shots);
(ii) shot amplitudes and durations do not have typical values
but are smoothly distributed;  (iii) occurrence of flares
is nearly random.  
Power spectra of such light curves show a $1/f^\alpha$
decline (with $\alpha = 1-2$ in normal Seyferts)
at high frequencies.  NLS1s exhibit a more peaked
profile and thus $\alpha$ is smaller.

\begin{figure}[t]

\centerline
{\psfig{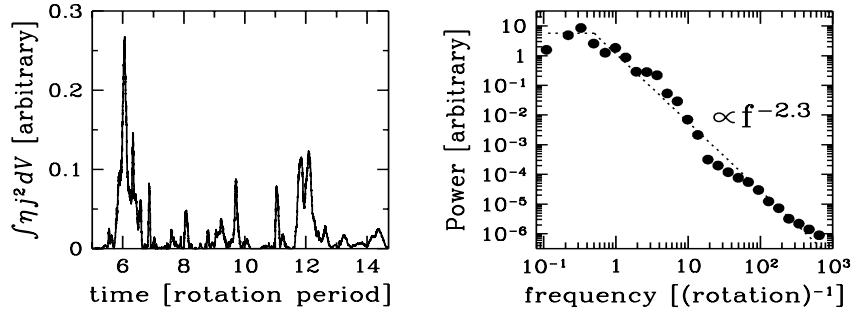}}
\caption
{A typical light curve (left) and its power-spectral density (right)
of the simulated MHD disk.
Here, we assume that the radiation is
predominantly due to field dissipation by magnetic reconnection;
thus we plot the temporal variation of Ohmic dissipation
integrated over the whole disk (see \cite{Kawa}).
}
\end{figure}

It is often suggested that variability
is due to magnetic reconnection leading to flares, as is the case in solar flares.
To investigate this magnetic-flare model,
we have examined the 3D MHD simulation data \cite{Machi}, finding
(i) spatial fractals in the $j/\rho$ distribution
(where $j$ and $\rho$ are current and matter densities, respectively) 
and
(ii) temporal $1/f^\alpha$ fluctuations in
the Ohmic dissipation, $\int\eta j^2 dV$
(where $\eta$ is resistivity, see Figure 3).
These two are closely related.
When reconnection occurs in fractal magnetic fields,
a variety of flare amplitudes arise, yielding
$1/f^\alpha$ fluctuations 
(see \cite{Kawa} for a more detailed discussion).

If fluctuations are of a magnetic origin,
large-amplitude fluctuations indicate a relatively large 
field energy compared with the radiation energy.
In the standard-type disk, cooling is efficient.
Since the emitted radiation energy 
is comparable to the gravitational-energy release, 
the internal energy of the gas should be much less than 
the gravitational energy, which the magnetic field energy
cannot exceed.  Hence, the magnetic field energy is 
much less than the emitted radiation energy, leading to
rather small fluctuations.

In the slim-disk case, on the other hand,
radiative cooling is inefficient.  
It is the trapped photons that contain large energy
comparable to gravitational energy, which
greatly exceeds the emitted radiation energy.
Magnetic energy can grow to overcome the emitted radiation energy.
Thus, large fluctuations are inevitable in radiation from
slim disks, consistent with the observations of NLS1s.
A more detailed discussion is presented elsewhere \cite{Mine}.

\end{document}